# Tetrahedral entropy captures the non-monotonicity of electrical conductivity in aqueous monatomic ions

Puja Banerjee and Biman Bagchi*

Solid State and Structural Chemistry Unit, Indian Institute of Science, Bangalore, India-560012

Corresponding author's E-mail: *profbiman@gmail.com/ bbagchi@iisc.ac.in*

## *Abstract*

The intriguing relationship between entropy and diffusion is a subject of much current interest. However, the experimentally observed unusual non-monotonic dependence of limiting ionic conductivity on inverse ion size is neither described by the Adam-Gibbs entropy crisis theory nor by the Rosenfeld entropy scaling. This failure is obvious because throughout the size variation the bulk entropy of the solvent remains the same, or undergoes infinitesimal change. We show that *it is the entropy experienced by the tagged ion that needs to be calculated*. This entropy can be quantified, at least partly, by the change in the tetrahedral ordering of water molecules in the hydration layer of the ions which exhibits a nonmonotonic size dependence.

## I. Introduction

The dynamics of monatomic alkali cations and halide anions in polar medium has been a subject of huge research interest over the past half-century[1-7]. Experimental results of the conductivities of these ions have long suggested a breakdown of the Stokes-Einstein's hydrodynamic relation[8,9]. This breakdown is popularly known as the "breakdown of Walden's product" (which is the product of the limiting ionic conductivity of the ion, $\Lambda_0$, and solvent viscosity $\eta_0$). To explain this deviation, several different explanations have been suggested[3,5,10].

In recent years, many studies have addressed the correlation between diffusion and entropy. The two foremost relations are those of Rosenfeld [11] and Adam-Gibbs [12]. Rosenfeld relation discussed the connection between diffusion and excess entropy of the system.

$$D = a \exp(b S_{ex}) \tag{1}$$

Here a and b are the empirical constants. $S_{ex}$ is the excess entropy, defined with respect to the ideal gas entropy per particle, $S_{id}$ as

$$S_{ex} = S - S_{id} \tag{2}$$

$S_{ex}$ is clearly negative. As the entropy S decreases with lowering temperature, $S_{ex}$ decreases (becomes more negative) and diffusion decreases. The second popular relation between diffusion and entropy is the Adam-Gibbs relation given by

$$D = A \exp\left(-\frac{C}{T S_C}\right) \tag{3}$$

where $S_C$ is the configuration entropy, and A and C are the empirical constants.

Several earlier studies have investigated the excess entropy ($S_{ex}$) scaling of diffusivity for water and water-like liquids like $BeF_2$, $SiO_2$[13-17]. Chakravarty[18] have shown that excess entropy scaling does not obey for tetrahedral systems like water, silica etc. at moderate pressure dense liquid state points. We note that only an approximate expression was used in the earlier studies. In a nutshell, these studies used only a radial measure of order but neglected the role or effects of the orientational correlations. In a recent perspective article Dyre has discussed excess entropy scaling in detail[19].

This excess entropy term arises due to correlations and can vary depending on the thermodynamic conditions. At higher density and lower temperature, excess entropy further decreases (becomes more negative) as correlation increases.

As $S_{ex}$ has its origin in the structure of the liquid, it can be defined as an expansion of multiparticle correlations

$$S_{ex} = S_2 + S_3 + S_4 + \cdots \tag{4}$$

Here $S_n(n=2,3,\ldots)$ is the excess entropy due to $n$-particle correlations. The translational part of the excess entropy is often approximated to two-body excess entropy, $S_2$ ($S_{ex}^{tr} \simeq S_2$) which can be expressed in terms of atom-atom radial distribution function. We directly write the expression in a binary liquid[20-23] where $g_{\alpha\beta}(r)$ is the radial pair correlation function

$$S_{\alpha\beta} = \int_0^\infty dr \left\{ g_{\alpha\beta}(r) \ln g_{\alpha\beta}(r) - \left[ g_{\alpha\beta}(r) - 1 \right] \right\} r^2 \tag{5}$$

Then the overall pair correlation entropy is calculated as

$$S_2 / Nk_B = -2\pi\rho \sum_{\alpha,\beta} x_\alpha x_\beta S_{\alpha\beta} \tag{6}$$

where N is the number of particles and $x_{\alpha/\beta}$ is the mole fraction of component α/β in the system. For one component liquid $S_2$ is observed to be a good approximation to $S_{ex}$[23].

Most of the earlier discussions focused on the variation of diffusion when entropy changes with temperature or pressure and the system is left unchanged. Here we have an interesting situation where diffusion changes substantially while the temperature and pressure remains unchanged but the solute-solvent system is changed. In this case, the most interesting question to ask is whether Adam-Gibbs or Rosenfeld relation is valid to explain the non-monotonic diffusivity of monatomic alkali cations in water at room temperature.

The above discussion requires a generalization for polyatomic molecules like water. In such cases, the entropy must contain the contribution from its rotational correlation. We now discuss such a generalized theoretical prescription.

If we neglect the vibrational contribution, the total entropy of water can be decomposed as

$$S = S_{id}^{tr} + S_{id}^{rot} + S_{ex}\left(\{\rho(r,\Omega)\},T\right) \qquad (7)$$

$S_{ex}$ derives its contribution from both translational and rotational motion. For an ideal system, the translational entropy, $S_{id}^{tr}$ is given by Sackur-Tetrode equation[24]

$$S_{id}^{tr} = \frac{5}{2}R + R\ln\left[\frac{V}{N}\left(\frac{2\pi m k_B T}{h^2}\right)^{3/2}\right] \qquad (8)$$

And the rotational entropy of ideal gas system can be obtained as[24]

$$S_{id}^{rot} = \frac{3}{2}R + R\ln\frac{1}{\pi\sigma}\left(\frac{8\pi^3 (I_A I_B I_C)^{1/3} k_B T}{h^2}\right)^{3/2} \qquad (9)$$

Here $I_A$, $I_B$ and $I_C$ are the components of moments of inertia along three principle axes. For a symmetric molecule rotational entropy is the highest as $I_A=I_B=I_C$. Here $\sigma$ is the symmetric factor, for an asymmetric molecule it is 1, for $H_2O$ it is 2 etc. In entropic unit (E.U.) at 1 atm pressure and 298K temperature ideal translational entropy of water, $S_{id}^{tr}$ is 17.41 and ideal rotational entropy, $S_{id}^{rot}$ for water is 5.45, which is almost 30% of the translational contribution.

Therefore, for a liquid like water, orientational ordering plays an important role to decide the total entropy of the system because depending on the nature of solute, structural ordering of water changes significantly. Kumar et al.[25] have used tetrahedral order parameter, $q_{tet}$ as a variable to define the contribution of excess entropy that comes from the ordering of water molecules.

$$S_{q_{tet}} = S_0 + \frac{3}{2}k_B \int_{q_{tet,min}}^{q_{tet,max}} \ln(1-q_{tet})P(q_{tet})dq_{tet} \qquad (10)$$

where $k_B$ is the Boltzmann constant and

$$S_0 = k_B \left[ \ln \Omega_0 + \frac{3}{2}\ln\frac{8}{3} \right]. \qquad (11)$$

The orientational tetrahedral order parameter[26, 27], $q_{tet}$ can capture the change in the local tetrahedral network of water molecule in the presence of solutes and is defined as

$$q_{tet} = 1 - \frac{3}{8}\sum_{j=1}^{3}\sum_{k=j+1}^{4}\left(\cos\psi_{jk} + \frac{1}{3}\right)^2 \qquad (12)$$

where $\psi_{jk}$ is the angle between the bond vectors $r_{ij}$ and $r_{ik}$ where j and k labels four nearest neighbour oxygen atoms of an oxygen atom of water molecules. For an ideal tetrahedral structure, this order parameter becomes 1 and for ideal gas it is zero. Sometimes, not the overall tetrahedrality of a system is important, but the local tetrahedrality that explain the effect. In this article we have discussed it in detail. The local tetrahedrality of an water-oxygen atom 'i' is defined as

$$q_{tet,i} = 1 - \frac{3}{8}\sum_{j=1}^{3}\sum_{k=j+1}^{4}\left(\cos\psi_{jik} + \frac{1}{3}\right)^2 \qquad (13)$$

Kumar et al.[25] considered local tetrahedral order of each molecules, $q_{tet,i}$ (i=1,2,...,N) and calculated the number of states in the range $(q_{tet,1}, q_{tet,2},...,q_{tet,N})$ and $(q_{tet,1}+\Delta q_{tet,1}, q_{tet,2}+\Delta q_{tet,2},...,q_{tet,N}+\Delta q_{tet,N})$ which are distributed in a hypercubic space. According to Eq. (13), $\frac{8}{3}(1-q_{tet,k})$ = constant defines a six-dimensional hypersurface with six tetrahedral angles, $\Psi_{jik}$. The approximation that the order parameter of each molecules, $q_{tet,i}$ are independent to each other gives an expression for the number of states

$$\Omega(q_{tet,1}, q_{tet,2},...,q_{tet,N}) \equiv \Omega_0^N \prod_{k=1}^{N}\left[\frac{8}{3}(1-q_{tet,k})\right]^{3/2} \qquad (14)$$

This defines the tetrahedral entropy as the logarithm of number of states of the system as

$$S(q_{tet,1}, q_{tet,2},...,q_{tet,N}) \equiv NS_0 + \frac{3}{2}k_B\sum_{k=1}^{N}\ln(1-q_{tet,k}) \qquad (15)$$

where $\Omega_0$ is a constant and so as $S_0$ (Eq. (11)). Now, if we have a distribution of tetrahedral order parameter, $P(q_{tet})$, Eq. (15) can be rewritten as Eq. (10).

It is important to note here is that the variation of the solvent radial distribution function, g(r) and orientational ordering, $q_{tet}(r)$ are independent to each other. That is the reason why these two contributions of pair correlation entropy, $S_2$ and tetrahedral entropy ($S_{q_{tet}}$) can be summed up to get the effective excess entropy of a system with orientational ordering.

In this article, we focus to verify the excess entropy scaling relation for a charged solute in water. We have carried out atomistic simulations of a number of systems with chloride salt solutions with different alkali cations. At first, we have checked if the pair-correlation entropy can explain the non-monotonous behavior of alkali cations in water, then we have investigated the orientational ordering of water around these ions and quantified the contribution of it to the excess entropy.

## II. Simulation details

Molecular dynamics simulations of five chloride salts, lithium chloride(LiCl), sodium chloride (NaCl), potassium chloride (KCl), rubidium chloride (RbCl) and cesium chloride (CsCl) in water have been carried out using the Lammps package[28]. Rigid non-polarizable force field parameters have been used for water as well as ions. SPC/E model[29] has been employed for water. For ions, potential parameters from Ref. [4] have been employed. The self-interaction parameters are listed in Table 1 and consist of Lennard-Jones and Coulombic terms.

We have taken 16 ion pairs in 8756 number of water molecules which corresponds to a concentration of 0.1M. The long-range forces were computed with Ewald summation[30, 31]. Trajectory was propagated using a velocity Verlet integrator with a time step of 1 fs. The aqueous salt solutions were equilibrated for 300 ps at 300 K and then a 2 ns MD trajectory was generated in the microcannonical (NVE) ensemble. The coordinates were stored every 5 fs for subsequent use for the evaluation of various properties.

*Table 1: Values of Lennard-Jones and electrostatic interaction potential parameters. e represents the magnitude of the electronic charge.*

| Atom, i | $\sigma_{ii}$(Å) | $\varepsilon_{ii}$(kJ/mol) | $q_i$ (e) | Ref |
|---|---|---|---|---|
| $H^w$ | 0.000 | 0.000 | +0.4238 | 29 |
| $O^w$ | 3.169 | 0.6502 | -0.8476 | 29 |
| $Li^+$ | 1.505 | 0.6904 | +1.0 | 4 |
| $Na^+$ | 2.583 | 0.4184 | +1.0 | 4 |
| $K^+$ | 3.331 | 0.4184 | +1.0 | 4 |
| $Rb^+$ | 3.527 | 0.4184 | +1.0 | 4 |
| $Cs^+$ | 3.883 | 0.4184 | +1.0 | 4 |
| $Cl^-$ | 4.401 | 0.4184 | -1.0 | 4 |

## III. Results and discussion

### A. Excess entropy from pair-correlation

We have extracted radial distribution function, g(r) between ion and water from the simulation data (shown in Figure 1(a)) which agrees well with the previous studies[32, 33]. We have considered pair correlation function, $g_{\alpha\beta}(r)$ between all the pairs of atoms including cation/anion-water oxygen, cation/anion-water hydrogen, cation-anion, water oxygen-water oxygen etc. and computed the total pair-correlation entropy using Eq. (5) and (6). It gives a very small variation of the value of $S_2$ (values are given in **Table 2**).

Table 2: Values of pair correlation entropy for different salt solutions considering all the pair interactions.

| System | $S_2/Nk_B$ |
|---|---|
| LiCl-water | -21.734 |
| NaCl-water | -21.606 |
| KCl-water | -21.736 |
| CsCl-water | -21.9448 |

Now, in the above analysis, we have considered all the pair interactions and calculated the total pair-correlation entropy using Eq. (6). As we have already mentioned, to obtain the conductivity/ diffusivity trend of the monatomic ions, the desired quantity is the excess entropy experienced by the ions. Therefore, the contributions of primary importance are: (a) the ion-water pair correlation entropy, $S_{cation-wat}$ and (b) orientational entropy in the first and second solvation shell. In case of a very dilute solution, the contribution from ion-water pair correlation is negligible. Only local structure around an ion undergoes a significant change while the rest of the bulk water remains unperturbed. It is thus accurate as we show below that is the local tetrahedral ordering that captures the nonmonotonicity.

This is further given in Table 3 that the radial part of the pair correlation between ion-water fails to capture the nonmonotonicity. Figure 1(a) shows the radial distribution function of water oxygen around cations and Figure 1(b) is the integrand of Eq. (5) for the ion-water oxygen pairs which shows a gradual decrease in peak heights. We have tabulated the values of $S_2$ in Table 3 which are obtained by considering only radial distribution function between cation-water oxygen atoms. The magnitude of this pair-correlation entropy decreases monotonically as the ion-size increases. Therefore, only the radial arrangement of water molecules that is included in the $g_{cation-water}(r)$ is not responsible for the non-monotonic nature of conductivity of these cations but the orientation that matters. We need the function, $g_{cation-water}(r, \Omega)$ where $\Omega$ denotes the orientation of water molecules around a cation.

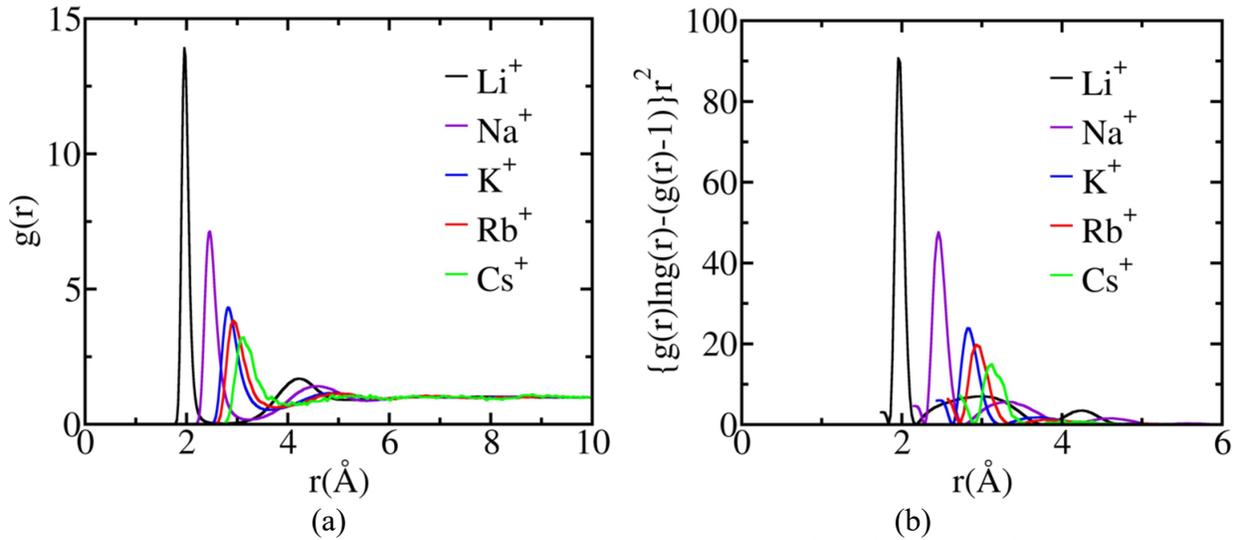

Figure 1:(a)Radial distribution function between alkali cations ($Li^+$, $Na^+$, $K^+$, $Rb^+$, $Cs^+$) and water oxygen atoms, (b) the monotonic nature of the integrand of pair-correlation entropy.

Table 3: Values of pair correlation entropy contribution arises from ion-water oxygen interaction.

| System | $S_2/Nk_B$ |
|---|---|
| $Li^+$-Ow | -0.239 |
| $Na^+$-Ow | -0.171 |
| $K^+$-Ow | -0.092 |
| $Rb^+$-Ow | -0.08 |
| $Cs^+$-Ow | -0.069 |

## B. Local tetrahedral ordering of water around ions

Tetrahedral order parameter of water (defined in Eq. (13)) can indeed capture the orientational ordering of water around an charged species. The distribution of tetrahedral order parameter, $q_{tet}$ of bulk water in five salt solutions are calculated using Eq. (13) and are shown in Figure 2. This is observed to be similar to the system of pure water as in a dilute solution of 0.1 M the overall ordering of water in the system does not get affected significantly.

The distribution mainly shows two characteristic peaks, one at $q_{tet}$~0.75 ad another at $q_{tet}$~0.5. The first one corresponds to the water molecules with four neighbors arranged in a tetrahedral structure and the second one represents non-tetrahedrally arranged water molecules. $q_{tet}$~0.5 signifies the local coordination number, n=3.[15]

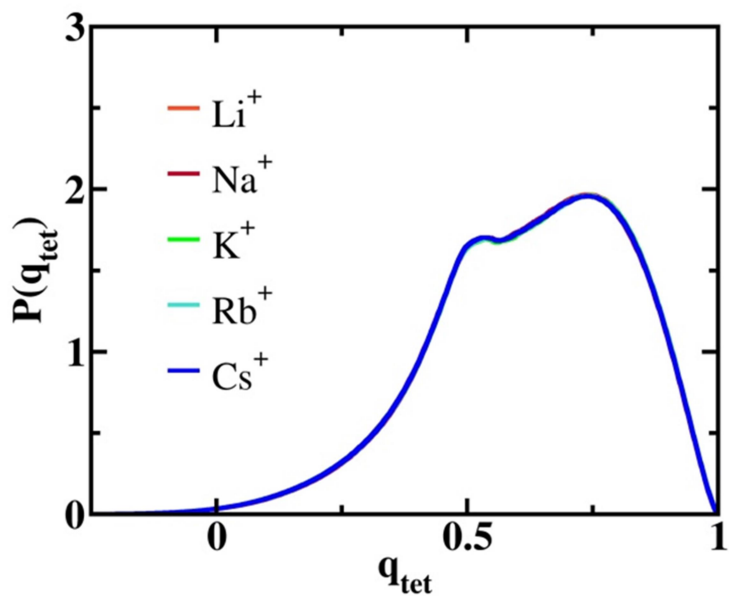

*Figure 2: Distribution of tetrahedral order parameter, $q_{tet}$ of bulk water molecules in different ionic solutions at 300 K.*

We next analyze the local tetrahedral ordering of water molecules in the first solvation shell of different alkali cations using Eq. (13). Figure 3(a) shows that the distributions are quite different especially for $Li^+$. Other than that, the population of $q_{tet} \sim 0.75$ differs significantly for different ions. Now, using these distributions, we calculate tetrahedral entropy using Eq. (10) and the values are shown in Figure 3(b).

The basic idea of tetrahedral entropy, $S(q_{tet})$ ( Eq. (10)) is that the structure that resembles to more and more like ideal tetrahedral ($q_{tet}=1$), contributes more to the magnitude of excess entropy and leads to a lower total entropy of the system. Now, in bulk water ideal tetrahedral structure is modified to a distorted one with $q_{tet} \sim 0.75$. Figure 3(a) shows that for the hydration layer water molecules around $Cs^+$ ion, the population of tetrahedral structures water molecules is highest and for $Li^+$ it is the lowest. Also, for the hydration shell water molecules of $Li^+$, a pronounced population is observed at $q_{tet} \sim 0.5$.

These two together result in the non-monotonic behavior of the tetrahedral entropy of these ions. The higher magnitude of $S(q_{tet})$ signifies the lower total entropy. Although it captures a non-monotonic behavior, the relative contributions of $Li^+$ and $Cs^+/Rb^+$ does not truly verify excess entropy scaling to explain experimental diffusivity/conductivity of these ions.

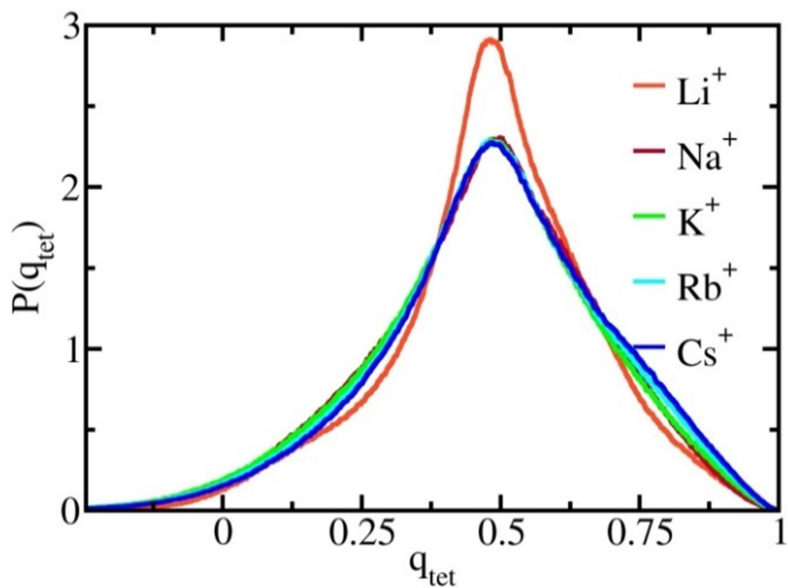

(a)

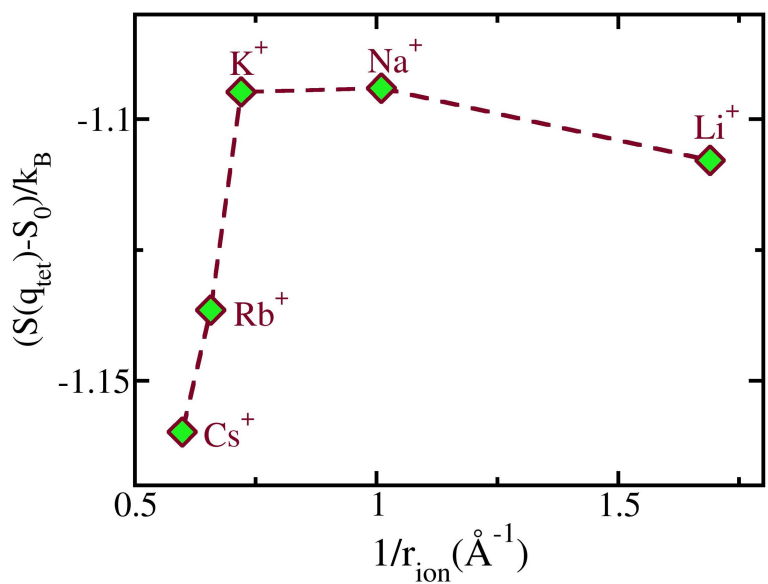

(b)

*Figure 3:(a) Comparison of the distribution of local tetrahedral order parameter, $q_{tet,i}$ of water molecules in the first hydration layer of different ions at 300 K, (b) comparison of the tetrahedral entropy, $S(q_{tet})$ of hydration water around different alkali cations.*

## C. Comparison of the structure and dynamics of positive and negative charges in water

We have computed the distribution of the local tetrahedral ordering in the first solvation shell of $K^+$ and $Cl^-$ ions (Figure 4(a)). This suggests that the water molecules in the hydration layer of $K^+$ with higher population of $q_{tet}$~0.75 are more tetrahedrally structured compared to hydration shell water molecules of $Cl^-$. We calculated the contribution of the hydration layer water molecules to the excess orientational entropy of these two ions using Eq. (10) and obtained the value of -1.1 for $K^+$ and -0.91 for $Cl^-$. This explains the higher diffusivity of $Cl^-$ ion.

We again characterize the structure of water molecules around these ions by measuring the polarization ($\cos\theta$; where $\theta$ is defined as the angle between ion-water intermolecular axis and water dipole vector). Figure 4(b) shows the polarization profiles of water molecules around $K^+$ and $Cl^-$ ions. Being opposite to each other due to different Coulomb's interaction, they even differ in the nature. They both exhibit two major peaks. The value of polarization in the first hydration shell of $K^+$ is much higher while the first polarization peak around $Cl^-$ ion is broader with lower value of $\cos\theta$. This result verifies the previous result of differently structured solvation shell around these two ions which are of similar sizes.

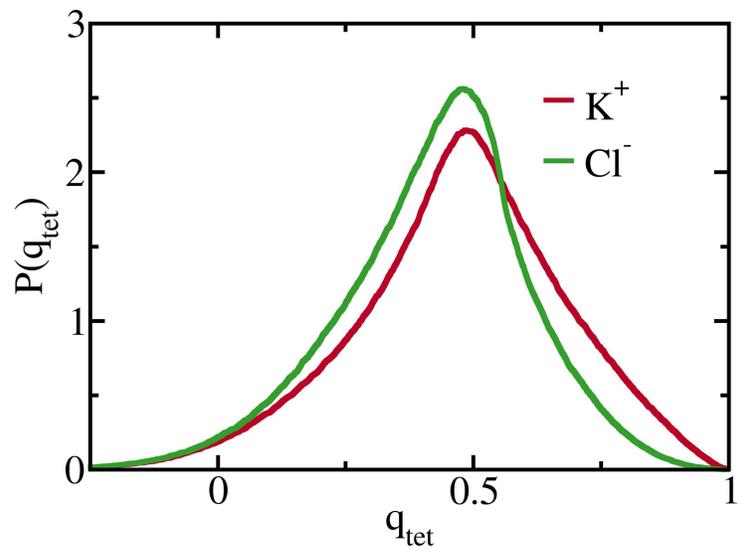

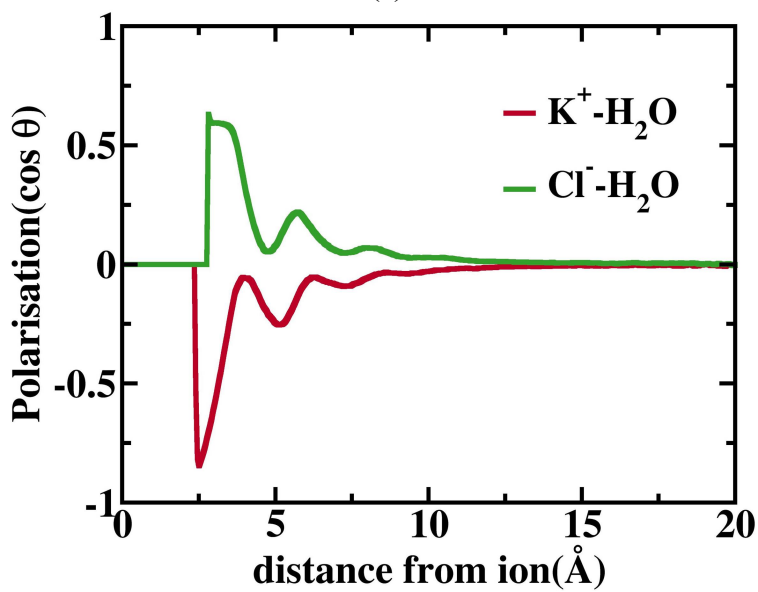

*Figure 4:(a) Comparison of the distribution of local tetrahedral order parameter, $q_{tet,i}$ of water molecules in the first hydration layer of positive ($K^+$) and negatively charged ($Cl^-$) ions at 300 K, (b) Polarisation profiles of water around isolated positive and negatively charged ions.*

## IV. Connection with earlier theories

The earliest attempt to explain the experimental observation of the non-monotonic size dependence of the limiting ionic conductivity or self-diffusion of singly positively charged ions was made by using a continuum model based approach[34-38]. Electrohydrodynamic equations were solved to obtain the excess friction due to the interaction between the charge of the ion and the dipolar solvent represented by a dielectric continuum with given static dielectric constant $\varepsilon_s$ and a Debye dielectric relaxation time, $\tau_D$. Thus, the continuum models ignore all aspects of solute-solvent molecularity except the size of the solute ion. The continuum model impressively reproduces the non-monotonic size dependence of diffusion but fails to capture the quantitative details.

Subsequent to the continuum model approach, a mode coupling theory was developed and used to explain the non-monotonic size dependence of the limiting ionic conductivity or self-diffusion of singly positively charged ions[3]. The MCT approach is nearly analytical. It requires as input the static structural correlation functions like ion-water dipole spatial and orientational correlation functions. Important new results of the MCT approach was the discovery of the importance of the ultrafast polarization relaxation modes, and a hidden role of the translational modes of the solvent molecules. A self-consistent solution of the MCT equation provide a semi-quantitative description of the size dependence of ions in water and methanol.

However, none of the above approaches make contact with the entropy theories of diffusion. Recently, however, Bhattacharyya and coworkers carried out an entropic view from the MCT approach for non-polar spherical solutes[39, 40].

## V. Conclusions

In this study, we have explored different contributions to the excess entropy in the systems of aqueous salt solutions. To obey Rosenfeld relation, the excess entropy should possess a non monotonic nature to explain the anomalous diffusivity of alkali cations in water. We have verified that only the pair-correlation entropy, $S_2$ cannot capture the expected trend.

As the rotational entropy of water contributes significantly to the total entropy, we have investigated the orientational ordering of water molecules in the hydration layer and observed that the entropy arises from the tetrahedral ordering of the water molecules could reproduce the expected non-monotonic size dependence of excess entropy.

We have also shown that the structural ordering of water around two similar sized but oppositely charged ions can be significantly different by computing distribution of local tetrahedral order parameter of hydration water and the polarization of water around these two ions. And these different structural ordering itself changes the entropy of the ions in water. The calculated tetrahedral entropy, $S(q_{tet})$ can explain the higher diffusivity of chloride ion compared to potassium ion.

## Acknowledgements

This work was supported in part by the Department ofScience and Technology (DST), Govt. of India, Sir J. C. Bosefellowship (to BB), and Council of Scientific and Industrial Research(CSIR), India(to PB).